\documentstyle[aps,twocolumn,prl,
graphicx,psfrag,amsmath,%
amsthm,amsfonts,amssymb,verbatim]{revtex}

\tightenlines

\newcounter{inta}
\newcounter{intb}
\newcounter{intc}
\begin{document}
\def\betaf{{$\beta{\rm eta}$}\ }
\newcommand{\Tr}{\mathop{\mathgroup\symoperators Tr}\nolimits} 
\newcommand{\Log}{\mathop{\mathgroup\symoperators Log}\nolimits} 
\hyphenation{mar-gi-nal}
\thispagestyle{plain}
\twocolumn[\hsize\textwidth\columnwidth\hsize\csname
@twocolumnfalse\endcsname
\title{On the Beta Function for Anisotropic Current Interactions in 2D}
\author{B. Gerganov$^a$, A. LeClair$^{a,b}$ and M. Moriconi$^a$%
%\thanks{\tt beg, leclair, moriconi @mail.lns.cornell.edu}
}
\address{~}
\address{$^a$Newman Laboratory, Cornell University, Ithaca, NY 14853}
\address{and}
\address{$^b$LPTHE, 4 Place Jussieu, Paris, France. }
\maketitle
\date{today}
\begin{abstract}
By making use of current-algebra Ward identities we study
renormalization of general anisotropic current-current interactions in
2D.  We obtain a set of algebraic conditions that ensure the
renormalizability of the theory to all orders.  In a certain minimal
prescription we compute the $\beta{\rm eta}$ function to all orders.
\end{abstract}
]
%

%\section{~}

Left-right current-current interactions in 2D arise in many physical
systems, for example in the Kosterlitz-Thouless transitions and in the
study of free electrons in random potentials. The chiral Gross-Neveu
model is the simplest example which is symmetry preserving
(isotropic)\cite{GN}.  Because such interactions are marginal in 2D,
the couplings are dimensionless and there is usually no small
parameter to expand in order to explore strong coupling.

In this work we determine the \betaf function to all orders in a
certain minimal prescription.  This should be sufficient for the study
of fixed points.  The primary tool that allows us to isolate the
$\log$ divergences at arbitrary order are the Ward identities for the
currents. Kutasov performed a similar computation in the simpler
isotropic case (i.e. non-abelian Thirring model, which is equivalent
to the chiral Gross-Neveu model) but argued his result was the leading
order in $1/k$ where $k$ is the level of the current
algebra\cite{Kutasov}. We believe our result to be exact.

We do not expect that there is anything dramatically new to learn for
the isotropic case. The behavior of models like the Gross-Neveu model
is very well understood. This is in contrast to the anisotropic,
i.e. symmetry breaking perturbations where one can expect richer
phenomena.  In the anisotropic case not all perturbations are in fact
renormalizable. We find three renormalizability conditions that ensure
the theory is renormalizable to all orders.  Given these conditions
are satisfied, we find a compact expression for the summation of all
orders in perturbation theory.

We know of no reason to expect additional non-perturbative corrections
to the \betaf function due for instance to instantons.  Our result
should be a useful tool for exploring strong coupling physics for the
many models in this class.  In this Letter we only report the main
result and defer applications to future publications.

%\bigskip

Consider a conformal field theory with Lie algebra symmetry
realized in the standard way \cite{KZ,WZW}. 
It possesses left and right
conserved currents, $J^a(z)$, $\bar{J}^a(\bar{z})$, where $z=x+iy$,
$\bar{z}=x-iy$, satisfying the operator product expansion (OPE)
\begin{equation}
   J^a(z)J^b(0) \sim \frac{k\eta^{ab}}{z^2} + 
	\frac{1}{z} f^{ab}_c J^c(0) + \ldots \ ,
\label{ope}
\end{equation}
and similarly for $\bar{J}^a(\bar{z})$. $\eta^{ab}$ in the above equation 
is a metric (Killing form) and $k$ is the level.

We include the case of Lie superalgebras with applications to 
disordered systems in mind. 
Here each current $J^a$ has a grade $[a] = 0$ or
$1$, and the tensors satisfy: $\eta^{ab} = {(-)}^{[a][b]} \eta^{ba}, 
f^{abc} = -{(-)}^{[b][c]} f^{acb}, f^{ab}_c = -{(-)}^{[a][b]} f^{ba}_c $,
where $f^{abc} = f^{ab}_i \eta^{ic}$.  For superalgebras, 
$\eta^{ab} $ is generally not diagonalizable, but we have
 $\eta_{ab}\eta^{bc} =
\delta_a^c$.

   The conformal field theory can be perturbed by marginal operators
which are bilinear in the currents. The most general action is
\begin{equation}
   S = S_{\rm cft} + \int 
\frac{d^2 x}{2\pi } \sum_{A} g_A {\cal O}^A \ , ~ {\cal O}^A = \sum_{a,\bar{a}}d^A_{a\bar{a}}J^a\bar{J}^{\bar{a}}
\ ,
\label{Gen:action}
\end{equation}
where $S_{\rm cft}$ is the conformal field theory with the
current-algebra symmetry, and 
$d^A_{a\bar{a}}$ are certain tensors that define the model. The
simplest case is that of a single coupling where $d_{a\bar{a}} =
\eta_{a\bar{a}}$ so that the perturbation is built on the Casimir and
preserves the symmetry.

For these models the \betaf function beyond 2 loops is prescription
dependent. Let $\beta_g = b_2 g^2 + b_3 g^3 + \ldots \ $.
The prescription dependence is equivalent to a redefinition of the
coupling $g'=g'(g)$. Let $g' = g + c g^2 + \ldots$ . One easily sees
that $\beta'(g') = b_2 {g'}^2 + b_3 {g'}^3+\ldots$, so that the one-
and two-loop coefficients $b_2$ and $b_3$ are universal. Though the
higher loop contributions are prescription dependent, the existence of
fixed points is not: ${\rm if} ~ \beta_g=0 \ , ~~ {\rm then} ~ \beta'(g') = 
\frac{\partial g'}{\partial g} \beta_g = 0 \ $.
For this reason, the particular prescription we will adopt is
meaningful.

We will compute the \betaf function in the following way. Let $X$ be
any arbitrary product of fields and suppose we isolate the 
ultra-violet logarithmic
divergences in the following way:
\begin{equation}
   \left\langle X \right\rangle 
	= \sum_{A} F_{A}(g,\log a)  
	\int\frac{d^2x}{2\pi} \left\langle {\cal O}^A(x) X \right\rangle 
+ ....\ .
\label{PT}
\end{equation}
where $a$ is a short distance cutoff (lattice spacing). 
In general $F_A$ has an expansion in powers of $\log a$:
\begin{equation}
   F_A = -a_A^{(0)}(g) + \sum_{n=1}^{\infty} a_A^{(n)}(g)\log^n a \ .
\label{a:exp}
\end{equation}
Requiring $d F_A / d\log a = 0$ gives
\begin{equation}
   \beta_{g_A} \frac{dg_A}{d\log a} = G^{-1}_{AB} a^{(1)}_B \ , ~~~ 
	G_{AB} = \partial_{g_B} a^{(0)}_A \ .
\label{beta:a1}
\end{equation}

   In our models equation (\ref{PT}) holds due to the Ward
identities for the currents.  This
makes the computation much simpler that the two-loop
computations in perturbed conformal field theory
carried out in \cite{Dotsenko}. 
 The order $g^n$ term in the expansion is
\begin{multline}
   \frac{{(-)}^n}{n!} g_{A_1} \ldots g_{A_n}
	d^{A_1}_{a_1\bar{a}_1} \ldots d^{A_n}_{a_n\bar{a}_n}
\\
   \times \int d1 \dots dn \left\langle 
	J^{a_1}\bar{J}^{\bar{a}_1}(1)
	\ldots J^{a_n}\bar{J}^{\bar{a}_n}(n) X
	\right\rangle ,
\label{O:n}
\end{multline}
where $\int d1 = \int d^2x_1 /2\pi$. The Ward identities read
\begin{multline}
   \left\langle J^{a_1}(1) \ldots J^{a_n}(n) X \right\rangle 
\\
   = \sum_{i \neq 1} \frac{k\eta^{a_1a_i}}{z^2_{1i}}
   \left\langle J^{a_2}(2) \ldots \widehat{J}^{a_i}(i) \ldots 
   J^{a_n}(n) X \right\rangle
\\
   + \frac{f^{a_1a_i}_c}{z_{1i}}
   \left\langle J^{a_2}(2) \ldots J^{c}(i) \ldots J^{a_n}(n) X 
   \right\rangle \ ,
\label{Ward}
\end{multline}
where $z_{ij} = z_i - z_j$, and $\widehat{J}$ means the current is
removed. (We do not display the factors of ${(-)}^{[a][b]}$ 
in the above equation and eq. (\ref{terms}) below.)
   In order to compute $F_A$ one only needs the integrals
\begin{eqnarray}
   \ ~~ \int \frac{d^2\rho}{2\pi} 
	\frac{1}{\rho-w}\frac{1}{\bar{\rho}-\bar{z}}
   & = & -\log|w-z| \ , ~~~~ (\theequation a)
\nonumber
\\
   \ ~~ \int \frac{d^2\rho}{2\pi} 
	\frac{1}{{(\rho-w)}^2}\frac{1}{\bar{\rho}-\bar{z}}
   & = & -\frac{1}{2}\frac{1}{w-z} \ , ~~~~~~~~ (\theequation b)
\nonumber
\\
   \ ~~ \int \frac{d^2\rho}{2\pi} 
	\frac{1}{{(\rho-w)}^2}\frac{1}{{(\bar{\rho}-\bar{z})}^2}
   & = & \frac{\pi}{2}\delta^{(2)}(w-z) \ , ~~~~ (\theequation c)
\nonumber
\end{eqnarray}
\addtocounter{equation}{1}
\setcounter{inta}{\arabic{equation}}
\def\theinta{\arabic{inta}$a$}
\setcounter{intb}{\arabic{equation}}
\def\theintb{\arabic{intb}$b$}
\setcounter{intc}{\arabic{equation}}
\def\theintc{\arabic{intc}$c$}
\\
\noindent
to systematically reduce (\ref{O:n}) to one integral.
In performing this reduction, we throw away two kinds of
contributions: (i) vacuum bubbles proportional to the volume $\int d^2x$;
(ii) $\log a$ divergences that factorize into a lower order 
contribution to $a^{(1)}$ times a finite part  which is cancelled
by dividing by $G_{AB}$.  
The remaining $\log a$ divergences have a 
rather simple structure.
Consider doing the integral $\int d1$ in (\ref{O:n}) using
(\ref{Ward}). One  is left with 3 kinds of terms $[k^2]$,
$[kf]$, and $[f^2]$ terms:
\begin{eqnarray}
   [k^2] &=& \frac{\pi k^2}{2} \int d2 \dots dn \sum_{i,j}
	\delta^{(2)}(z_{ij}) \eta^{a_1a_i}\eta^{\bar{a}_1\bar{a}_j}
\nonumber\\[-4pt]
   \ && ~~~ \times	
   \left\langle \ldots \widehat{J} \ \bar{J}^{\bar{a}_i}(i) 
   \ldots J^{a_j} \ \widehat{\bar{J}}(j) \ldots X 
   \right\rangle
\nonumber\\[2pt] \ 
   [kf] &=& -\frac{k}{2} \int d2 \dots dn \sum_{i,j}
\Big\{ \eta^{a_1a_i} f^{\bar{a}_1\bar{a}_j}_{\bar{c}}\frac{1}{z_{ij}}
\nonumber\\[-4pt] 
   \ && ~~~ \times	
   \left\langle \ldots \widehat{J} \ \bar{J}^{\bar{a}_i}(i) 
   \ldots J^{a_j} \bar{J}^{\bar{c}}(j) \ldots X 
   \right\rangle + z\rightarrow\bar{z} \Big\}
\nonumber
\end{eqnarray}
\begin{eqnarray}
  [f^2] &=& -\int d2 \dots dn \sum_{i,j}
	\log|z_{ij}| \> f^{a_1a_i}_c  f^{\bar{a}_1\bar{a}_j}_{\bar{c}}
\nonumber\\
   \ && ~~~ \times		
   \left\langle \ldots J^c \bar{J}^{\bar{a}_i}(i) 
   \ldots J^{a_j} \bar{J}^{\bar{c}}(j) \ldots X 
   \right\rangle
\label{terms} 
\end{eqnarray}

\noindent 
$\Log a$ divergences arise in the following ways: (i) The $[f^2]$ term
gives a $\log$ divergence when the remaining integrals give
$\delta^{(2)}(z_{ij})$ by repeatedly using (\theintc).  ($\log |z_{ij}
| = \log a $ when $i=j$, (ii) Repeatedly using (\theintb) in the
$[kf]$ terms can flip $1/z$ to $1/\bar{z}$ or vice versa a number of
times. Ultimately, a Ward identity involving another $f$ will give a
$\log(z_{ij})$ like in the $f^2$ term. Then again, the remaining
integrations must be proportional to $\delta^{(2)}(z_{ij})$ to give a
$\log a$ divergence.

   Integrations of $\log(z_{ij})$ with other integrands which are not
$\delta$-functions give rise to higher powers of $\log a$ but are not
important for the $\beta$-function. We checked explicitly at two loops
that the $\log^2\hspace{-0.5mm} a$ divergences are squares of one-loop
divergences.

   The above contributions can be represented graphically, which will
be useful for determining their tensor structure. Let solid lines
denote the $J(z)$ OPEs and dashed lines denote the
$\bar{J}(\bar{z})$ ones. Denote the $f$-OPEs with arrows.
For the holomorphic OPEs for instance:
\begin{figure}[htb] 
\begin{center}
\hspace{-15mm}
\psfrag{A}{$\sim ~~ k \eta^{a_i a_j}$}
\psfrag{B}{$\sim ~~ f^{a_i a_j}_c$}
\psfrag{c}{$\scriptstyle a_i$}
\psfrag{d}{$\scriptstyle a_j$}
\psfrag{e}{$\scriptstyle i$}
\psfrag{f}{$\scriptstyle j$}
%\psfrag{g}{$\scriptstyle \bar{a}_i$}
%\psfrag{h}{$\scriptstyle \bar{a}_j$}
%\psfrag{i}{$\scriptstyle i$}
%\psfrag{j}{$\scriptstyle j$}
\psfrag{k}{$\scriptstyle c$}
\psfrag{l}{$\scriptstyle \bar{c}$}
\includegraphics[width=3.4cm]{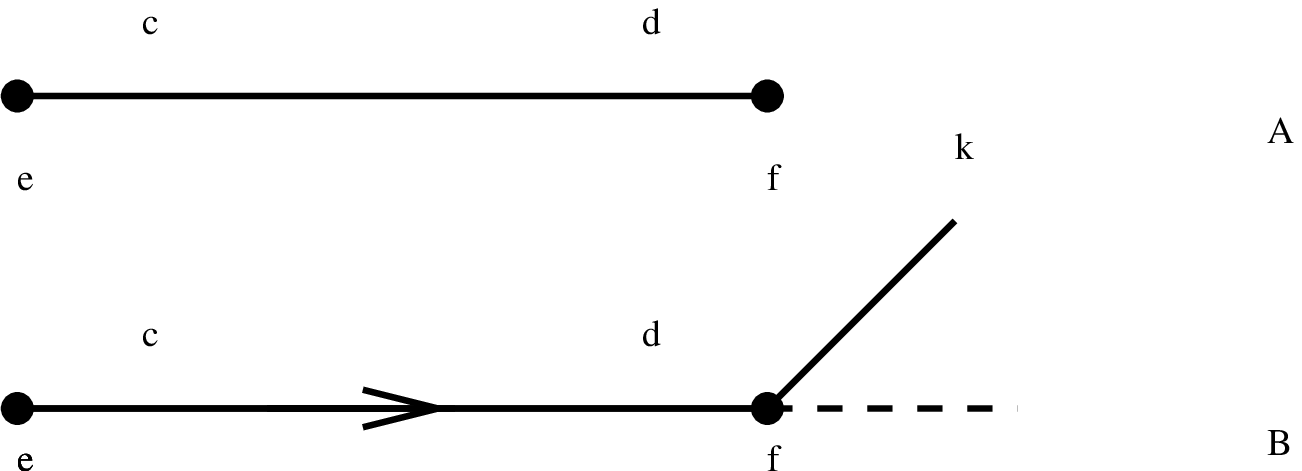} 
\end{center}
\vspace{-2mm}
\caption{Diagramatic representation of $k$- and $f$-OPEs.} 
\vspace{-2mm}
\label{Figure1} 
\end{figure} 
\noindent

   First consider $n={\rm even}$, which corresponds to an odd number
of loops. There are two types of contributions: Type 2A, 2B. Type 2A
is shown graphically in Figure \ref{Figure3}:
\begin{figure}[htb] 
\begin{center}
\psfrag{a}{$\scriptstyle 1$}
\psfrag{b}{$\scriptstyle 2$}
\psfrag{c}{$\scriptstyle 3$}
\psfrag{d}{$\scriptstyle m-1$}
\psfrag{e}{$\scriptstyle m$}
\psfrag{f}{$\scriptstyle m+1$}
\psfrag{g}{$\scriptstyle m+2$}
\psfrag{h}{$\scriptstyle n$}
\includegraphics[width=5.1cm]{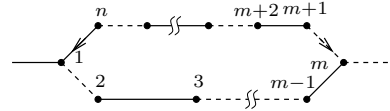} 
\end{center}
\vspace{-2mm}
\caption{Type 2A diagram.} 
\vspace{-2mm}
\label{Figure3} 
\end{figure} 
\noindent
The diagrams represent successive integrations from top to bottom. The
top alternating holomorphic and anti-holomorphic lines represent
succession of $1/z $ to $1/\bar{z}$ flips using (\theintb). The bottom
alternating lines represent $\delta$-functions coming from
(\theintc). The rest of the diagram represents the $f^{ab}_c
f^{\bar{a}\bar{b}}_{\bar{c}}$ leading to the $\log$ divergence coming
from (\theinta). The external lines represent the leftover
$J\bar{J}$ in (\ref{PT}). In the Type 2A diagrams, $m$ is odd,
$m=1,3,\dots,n-1$, and there are $n/2$ diagrams of this type. The Type
2B diagrams are distinguished by having the current that remains after
the $f^{ab}_c$ OPE becoming an internal part of the diagram:
\begin{figure}[htb] 
\begin{center}
\psfrag{a}{$\scriptstyle 1$}
\psfrag{b}{$\scriptstyle 2$}
\psfrag{c}{$\scriptstyle 3$}
\psfrag{d}{$\scriptstyle m-1$}
\psfrag{e}{$\scriptstyle m$}
\psfrag{f}{$\scriptstyle m+1$}
\psfrag{g}{$\scriptstyle m+2$}
\psfrag{h}{$\scriptstyle n$}
\includegraphics[width=5.1cm]{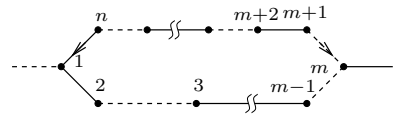} 
\caption{Type 2B diagram.} 
\end{center}
\label{Figure4}
\end{figure} 
\vspace{-2mm}
\noindent 
Again $m$ is odd, $m=3,5,\dots,n-1$. There are $(n-2)/2$
diagrams of this type.

   At $n={\rm odd}$ order, diagrams involve two holomorphic $f$'s or two
anti-holomorphic $f$'s. There are again 2 types, 1A, 1B. The 1A
diagrams are
\begin{figure}[htb]
\begin{center} 
\psfrag{a}{$\scriptstyle 1$}
\psfrag{b}{$\scriptstyle 2$}
\psfrag{c}{$\scriptstyle 3$}
\psfrag{d}{$\scriptstyle m-1$}
\psfrag{e}{$\scriptstyle m$}
\psfrag{f}{$\scriptstyle m+1$}
\psfrag{g}{$\scriptstyle m+2$}
\psfrag{h}{$\scriptstyle n$}
\includegraphics[width=5.1cm]{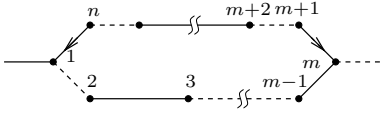} 
\end{center}
\caption{Type 1A diagram.} 
\label{Figure5} 
\end{figure} 
\noindent
Here $m=$ odd $=3,5,\dots,n-2$. There are $(n-3)/2$ of these.
A Type 1B diagram is distinguished by both $f$'s coming to the same 
point:
\begin{figure}[htb]
\begin{center} 
\psfrag{a}{$\scriptstyle 1$}
\psfrag{b}{$\scriptstyle 2$}
\psfrag{c}{\hspace{2mm}$\scriptstyle n-2$}
\psfrag{d}{\hspace{2mm}$\scriptstyle n-1$}
\psfrag{e}{$\scriptstyle n$}
\includegraphics[width=3.4cm]{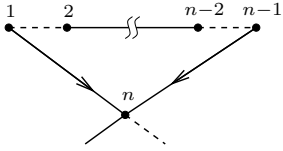} 
\end{center}
\caption{Type 1B diagram.} 
\label{Figure6} 
\end{figure} 
\noindent
There is one diagram of this type. It is the only contribution 
at two loops.   For $n$ odd each of the
above diagrams has a ``$z/\bar{z}$ relative'' where dashed lines are
replaced with solid and vice versa. These give the same contribution.
The total number of diagrams is then $(n-1)$ for  $n$ even or odd.   

   Each diagram represents a tensor built out of $\eta^{ab}$ and
$f^{ab}_c$ which must be contracted with $d^A_{ab}$'s.  An arbitrary
choice of $d^A_{ab}$ does not lead to a renormalizable theory,
i.e. (\ref{PT}) does not hold. At one loop one needs:
\begin{equation}
   {(-)}^{[b][c]} d^A_{ab} d^B_{cd} f^{ac}_i f^{bd}_j 
	= \sum_C C^{AB}_C d^C_{ij} 
\label{Cond1}
\end{equation}
for some structure constants $C^{AB}_C$. This closure of the operator
algebra of ${\cal O}^A$ is equivalent to the OPE
\begin{equation}
   {\cal O}^A(z,\bar{z}){\cal O}^B(0)
	\sim  \frac{1}{z\bar{z}} C^{AB}_C {\cal O}^C(0) \ 
\label{OPE:pert}
\end{equation}
and it is well-known that $C^{AB}_C$ determine the \betaf function to
one loop\cite{Zamo}.  This one-loop condition does not ensure
renormalizability at higher loops.  Systematically examining the
higher loop corrections we find that one also needs the quadratic form
$d^A$ to satisfy
\begin{eqnarray}
   \eta^{ij} d^A_{ai} d^B_{bj} 
	&=& \sum_C D^{AB}_C d^C_{ba} 
\\[-2pt]
   d^A_{ij} f^{ja}_k f^{ik}_b 
	&=& \sum_B R^A_B \eta^{ac} d^B_{cb} 
\ \label{Cond2}
\end{eqnarray}
The conditions (\ref{Cond1}) and (\ref{Cond2}) guarantee that the
theory is renormalizable to all orders. The conditions (\ref{Cond2})
already arise at 2 loops.  One can show from the defining relations
that $D, C $ satisfy $D^{AB}_C = D^{BA}_C, C^{AB}_C = C^{BA}_C, 
D^{AC}_D D^{DB}_E = D^{AB}_D D^{DC}_E$.

The structure constants $D,R$ also have a meaning in terms of
OPEs. Define $T^A(z)$ as
\begin{equation}
   T^A(z) = d^A_{ab} J^a (z)  J^b (z)  \ 
\label{T:A}
\end{equation}
suitably normal ordered.  $T^{A}$ is a kind of energy-mom-entum tensor
built out of $d^A_{ab}$ and the left-moving currents only.  In the
symmetric case $T^A$ is the affine-Sugawara stress-tensor up to a
normalization.  Then one finds
\begin{equation}
   T^A(z) {\cal O}^B(0) \sim \frac{1}{z^2} 
	\left( 2k D^{AB}_C  
+  R^A_E D^{BE}_C \right) {\cal O}^C(0)  
\ ,
\label{OPE:T-A}
\end{equation}
In practice this is the most efficient way to compute the RG data.
The $D$ term is distinguished from the $RD$ term by being proportional
to $k$.  As shown below, the \betaf function depends only on the
combination $RD$ so there is no need to compute $R$ separately. 
The conditions under which $T^A$ actually define a Virasoro
algebra\cite{Halpern} 
are stronger than our renormalizability conditions.  
 
   Finally we can write down the beta function. Let us arrange the
couplings into a row vector $g=(g_1,g_2,\ldots)$. It is convenient to
define a matrix of couplings $D(g)$, ${D(g)}^A_B = \sum_C D^{AC}_B \>
g_C$.  Then $g D^n(g)$ is also a row vector. The 4 kinds of
contributions (coming from the Type 2A through Type 1B diagrams
respectively) are
\begin{eqnarray}
   \  \beta^{\rm (2A)}_{g_A}  \hspace{-1.7pt} &=& \hspace{-1.7pt}
   -\frac{1}{2}  {\left( \frac{k}{2} \right)}^{n-2}
   \hspace{-1.7pt}	
	{\left( g D^{m-1} \right)}_B
	{\left( g D^{n-m-1} \right)}_C
	C^{BC}_A \ 
\nonumber\\[-2pt]
   \  \beta^{\rm (2B)}_{g_A}  \hspace{-1.7pt} &=& \hspace{-1.7pt}
   -\frac{1}{2} {\left( \frac{k}{2} \right)}^{n-2}
   \hspace{-1.7pt}	
	{\left( g D^{m-3} \right)}_B
	{\left( g D^{n-m-1} \right)}_C
	C^{BC}_E (D^2)^E_A \ 
\nonumber\\[-2pt]
   \  \beta^{\rm (1A)}_{g_A}
\hspace{-1.7pt} &=& \hspace{-1.7pt}
   \  {\left( \frac{k}{2} \right)}^{n-2}
   \hspace{-1.7pt}	
	{\left( g D^{m-2} \right)}_B
	{\left( g D^{n-m-1} \right)}_C
	C^{BC}_E D^E_A \ 
\nonumber\\[-2pt]
   \  \beta^{\rm (1B)}_{g_A} 
\hspace{-1.7pt} &=& \hspace{-1.7pt}
   - {\left( \frac{k}{2} \right)}^{n-2}
   \hspace{-1.7pt}	
	{\left( g D^{n-2} R D\right)}_A \ ,
\ \label{4types}
\end{eqnarray}
where $D = D(g)$.  (Each diagram comes with a symmetry factor
of $n!/2$.)   

   Remarkably, the series can be summed in a closed form.  To display
the result in a compact way, for any two vectors $v^1$ and $v^2$ let
$C(v^1,v^2)$ denote a new row vector, $C(v^1,v^2)_A = \sum_{B,C} v^1_B
v^2_C \> C^{BC}_A$.  Defining $g'= g \left( 1-k^2 D^2(g)/4
\right)^{-1}$, then
\begin{multline}
   \beta_g = -\frac{1}{2} C(g',g')\left( 1 + k^2 D^2/4 \right)
\\[-4pt]
   + \frac{k^3}{8} C(g'D,g'D) D - \frac{k}{2} g'DRD \ , 
\label{TheBeta}
\end{multline}
where again $D=D(g)$.  The above \betaf function determines
how couplings flow with increasing {\it length} scale $a$, 
where large $a$ corresponds to the low-energy limit. 

   The simplest possible case corresponds to a single coupling with
$d_{ab} = \eta_{ab}$. Normalizing the Lie algebra generators in the
fundamental representation $F$ so that $\Tr ( t^a t^b )$ $= c_F
\delta^{ab} = \eta^{ab}$ and using $\eta_{ij} f^{jc}_k f^{ik}_d =
C_{\rm adj} \delta^c_d$, one finds
%
%Let us normalize the Lie algebra generators in
%the fundamental representation $F$ so that $\Tr ( t^a t^b )$ $= c_F \delta^{ab}
%= \eta^{ab}$. Using $\eta_{ij} f^{jc}_k f^{ik}_d = C_{\rm adj}
%\delta^c_d$, one finds
%
\begin{equation}
   \beta_g = \frac{1}{2}  
	\frac{C_{\rm adj} g^2}{ {\left( 1 + kgc_F/2 \right)}^2 } \ .
\label{beta:Cas}
\end{equation}
This agrees with previous two-loop calculations for 
the Gross-Neveu model\cite{Two-Loop}. (The usual convention
corresponds to $c_F = 1/2$ with $k=1$.)  A three-loop
computation was performed in \cite{Three}.

From the result (\ref{beta:Cas}) we see that the \betaf function is
identically zero if $C_{\rm adj} = 0$.  This occurs for the
superalgebras $osp(2n+2|2n)$ and $PSL(n|n)$\cite{Bershad,Vafa,GLL}. In
the work\cite{GLL}, an all orders \betaf function was proposed in the
case of $gl(1|1)$ using a simple scaling argument, and the result had
a structure of the kind described here.
 
\def\CO{{\cal O}}
\def\zbar{\bar{z}}

   The renormalizability conditions (\ref{OPE:pert},\ref{OPE:T-A})
are rather restrictive. Consider for instance a completely anisotropic
perturbation $\sum_a g_a J^a\bar{J}^a = \sum_a g_a {\cal{O}}^a $
in a basis where $\eta^{ab}=\delta^{ab}$, i.e. the couplings $g_a,
g_b$ are not equal for $a\neq b$.  This is not renormalizable in
general since the operator algebra of ${\cal O}^a$ does not close:
\begin{equation}
   \CO^a (z,\bar{z}) \CO^b (0) \sim \frac{1}{z\zbar} \> 
	\sum_{c,d} f^{ab}_c f^{ab}_d  J^c  \bar{J}^d (0) 
\label{close} 
\end{equation}
The algebra only closes if $[t^a, t^b]$ is proportional to a {\it
single} generator, since then 
$f^{ab}_c f^{ab}_d \propto \delta_{cd}$ 
and one has $C^{ab}_c = (f^{ab}_c)^2$.  
Adopting a basis for $J^a$ based on the root
system of the Lie algebra, in order to build a renormalizable theory
one needs to include additional interactions of the form $(\alpha \cdot
H)(\alpha' \cdot \bar{H})$ where $\alpha, \alpha'$ are roots and
$H$ is in the Cartan basis.  The resulting theory is renormalizable
and one can then compute the RG data $C, D, R$ and the \betaf
function.  Interesting models which are still renormalizable can be
obtained by equating subsets of the couplings consistent with a global
sub-symmetry.

The simplest example is $su(2)$ since it only has one Cartan
generator. 
Let us illustrate our main result in this example.
Let
the currents be normalized as
\begin{multline}
   J_3 (z)  J^{\pm} (0)  \hspace{-1.1pt} \sim \hspace{-1.1pt} 
	\pm \frac{1}{z} J^{\pm} (0), ~~
   J^+ (z) J^- (0) \hspace{-1.1pt} \sim \hspace{-1.1pt} 
	\frac{k}{2}\frac{1}{z^2} + \frac{1}{z} J_3 (0),
\\
   J_3 (z)  J_3 (0)  \sim \frac{k}{2}\frac{1}{z^2} \ . 
~~~~~~~~~~~~~~
\label{su2:curr}
\end{multline}
This corresponds to $\eta^{ab}=\frac{1}{2}\delta^{ab}$, $C_{\rm
adj} = 2$, and $c_F = 1/2$.  Consider the interaction
\begin{equation}
  \sum_A  g_A {\cal O}^A = g_{\perp}\left( J^+ \bar{J}^- 
	+ J^- \bar{J}^+ \right) 
   + g_{\parallel} J_3\bar{J}_3 \ .
\label{su2:int}
\end{equation}
Using (\ref{OPE:pert}) and (\ref{OPE:T-A}) the RG data are:
$D^{\perp\perp}_{\perp}=D^{\parallel\parallel}_{\parallel}= 1/2$,
$C^{\perp\parallel}_{\perp}=C^{\parallel\perp}_{\perp}=-1$,
$C^{\perp\perp}_{\parallel}=-2$,
${(RD)}^{\perp\perp}_{\perp}={(RD)}^{\parallel\perp}_{\perp} = 1$,
${(RD)}^{\perp\parallel}_{\parallel} = 2$.
After some algebra one finds
\begin{eqnarray}
   \beta_{g_\perp} &=& \frac{g_\perp 
   	\left( g_\parallel - k g^2_\perp/4\right)}
	{\left( 1 - k^2g^2_\perp/16 \right) 
 	 \left( 1 + k g_\parallel/4 \right)} \ 
\nonumber\\
   \beta_{g_\parallel} &=& \frac{g^2_\perp 
   	{\left( 1 - k g_\parallel/4\right)}^2}
	{{\left( 1 - k^2g^2_\perp/16 \right)}^2} \ .
\ \label{beta:su2}
\end{eqnarray}
When $g_\perp = g_\parallel$ one recovers (\ref{beta:Cas}).

One can use these \betaf functions to study Kosterlitz-Thouless
flows\cite{Kosterlitz} at strong coupling, or the issue 
of symmetry restoration recently addressed in \cite{Konik}.
One observes flows to the isotropic line $g_\perp 
= g_\parallel$ in the regime $0<g_{\perp, \parallel} < 4/k$.
However there are other features which will be reported
elsewhere, along with applications  
of the above results to disordered fermions\cite{disorder}.  

We hope to publish a longer version of this paper with
more examples\cite{Next}.  

\section*{Acknowledgments}

We would like to thank M. Ameduri, P. Argyres, Z. Bassi, D. Bernard,
V. Dotsenko, and P. Wiegmann for discussions. This work is in part
supported by the NSF.

\vspace{.4cm}
\end{document}